\documentclass[aps,prl,preprint,groupedaddress]{revtex4-1}

\usepackage{amsmath}

\begin{document}
	
	
	\title{Resolving the observer reference class problem in cosmology}
	
	
	\author{Simon Friederich}
	\email[]{s.m.friederich@rug.nl}
	\affiliation{
		University of Groningen, University College Groningen, Hoendiepskade 23/24, NL-9718 BG Groningen and\\
		University of Groningen, Faculty of Philosophy, Oude Boteringestraat 52, NL-9712 GL Groningen, The Netherlands }
	
	
	\date{\today}
	
	\begin{abstract} 
		The assumption that we are \textit{typical} observers plays a core role in attempts to make multiverse theories empirically testable. A widely shared worry about this assumption is that it suffers from systematic ambiguity concerning the reference class of observers with respect to which typicality is assumed. As a way out, Srednicki and Hartle recommend that we empirically test typicality with respect to different candidate reference classes in analogy to how we test physical theories. Unfortunately, as this paper argues, this idea fails because typicality is not the kind of assumption that can be subjected to empirical tests. As an alternative, a \textit{background information constraint} on observer reference class choice is suggested according to which the observer reference class should be chosen such that it includes precisely those observers who one could possibly be, given one's assumed background information.
	\end{abstract}
	
	\pacs{01.70.+w, 98.80.Es}
	
	\maketitle
	
\section{Introduction}

String theory in combination with eternal inflation suggests that our universe may be one among many which together form the so-called \textit{landscape multiverse} \cite{boussopolchinski,kachru,susskind}: a vast collection of island universes produced by eternal inflation, where conditions differ radically between island universes, which realize different string vacua and, so, have different effective laws and parameters. Enhancing the empirical testability of multiverse scenarios such as the landscape multiverse---or making them testable in the first place---may be a crucial prerequisite for further progress in cosmology and high energy physics. The strategy that is widely considered the most promising is to regard multiverse theories as predicting those observations that \textit{typical} observers among those who exist according to these theories will make \cite{vilenkin,bostrom_book,elga_ind}. Following Vilenkin, the term \textit{principle of mediocrity} is widely used for this assumption.

A serious worry about typicality is that it suffers from a principled ambiguity concerning the observer reference class. (See \cite{aguirretegmark,weinstein,hartlesrednicki,azhar} for versions of this worry.) The present paper considers a response to this worry that derives from the work of Srednicki and Hartle \cite{hartlesrednicki_new}: that we should empirically \textit{test} the assumption that we are typical observers with respect to competing candidate reference classes in analogy to how we test physical theories. Unfortunately, as this paper will argue, typicality with respect to specific reference classes fails to be the kind of assumption with respect to which the concept of an empirical test makes sense. However, a non-arbitrary way of singling out a systematically preferred observer reference class turns out to exist, namely, to include in the reference class precisely those observers who one possibly \textit{could} be, given one's background information $D_0$ and assuming the physical theory $T$ at issue is correct. The paper concludes by outlining the repercussions of this \textit{background information constraint} (BIC) on observer reference class choice with respect to attempts of accounting for the values of parameters like the cosmological constant using multiverse theories.


\section{Typicality and its utility in testing multiverse theories}

Typicality principles are used to overcome two severe problems that arise in attempts to test multiverse theories. The first problem is that any multiverse theory $T$ on its own makes little to no empirical predictions if the multiverse that it entails is sufficiently vast and diverse. There may well exist countless different observers across universes with qualitatively identical empirical background information as ourselves who are bound to make radically different observations in their futures \cite{hartlesrednicki_new}. This makes it at least prima facie unclear which observations such a theory $T$ should be regarded as predicting for our future in view of our present and past.

The second problem is that we may want to account for (``postdict'') the values of parameters that we have already measured in our own universe using multiverse theories. But if according to some multiverse theory $T$ the values of parameters differ wildly between universes over some wide range, it is again at least prima facie unclear which measured values we should regard as accounted for by $T$. For example, according to the landscape multiverse scenario the value of the cosmological constant varies randomly across island universes. As famously pointed out by Weinberg \cite{weinberg}, only values in some restricted range are compatible with life, and it is unsurprising that the value that we find lies within that range. But the landscape multiverse entails the existence of universes with \textit{all} values in this ``anthropically allowed'' range (and many more outside), so it seems unclear for which measured values we should regard the landscape multiverse as confirmed and for which ones as disconfirmed.

The assumption that we are \textit{typical} observers can at least in principle help us overcome both these problems. Typicality straightforwardly follows from \textit{self-locating indifference} \cite{elga_ind}, \footnote{The term \textit{self-identifying indifference} would fit better, but \textit{self-locating indifference} is already widely used, so I stick to this terminology here.}: the principle that one should assign equal probability to being any specific observer in some specified observer reference class or, as put by Bostrom's \textit{self-sampling assumption} (SSA) \cite{bostrom_book}, that one should reason as if one were randomly sampled from the observer reference class. Applied to expectations concerning future observations, self-locating indifference recommends that we should expect to observe what most or ``typical'' observers in the reference class will observe. With respect to parameters that we have already measured, typicality advises us to regard cosmological theories $T$ as able to account for precisely those values that typical reference class members observe if $T$ is true. Virtually all suggested attempts to account for the measured value of the cosmological constant using the landscape multiverse scenario (e.g. \cite{boussoharnik,desimone}) are based on this idea.

As outlined above, a widely shared worry about typicality is that it suffers from a profound and unresolvable unclarity concerning observer reference class. This worry can be introduced and illustrated using an example due to Hartle and Srednicki \cite{hartlesrednicki}, \footnote{See (\cite{carroll_book}, p.\ 225), for an analogous example which posits intelligent lizard beings on a planet orbiting Tau Ceti rather than Jovians and (\cite{weinstein}, Sect.\ 2.2), for an example featuring humans and ``aliens''.}: imagine that we have an empirically well-motivated theory $T_J$ which entails that there are hitherto unknown intelligent observers in the atmosphere of Jupiter (``Jovians''), in fact are many more Jovians than there are humans on Earth, i.e. $N_J\ll N_H$ for their numbers; a alternative to $T_J$ we consider a ``null hypothesis'' $T_0$ according to which there are no Jovians but only $N_H$ humans. (It is assumed that there are no other candidate observers beside humans and Jovians that we need to consider.)

The problem of observer reference class choice becomes manifest in the question of whether we should include the Jovians in the reference class. The answer to this question may dramatically impact our preference among $T_J$ and $T_0$. Assume, for the sake of simplicity, that there are no observers outside our solar system, and suppose that we do include the Jovians in the reference class. Then, if $T_J$ is true, we are highly \textit{atypical} observers in the reference class because most reference class members are Jovians rather than humans. In contrast, if $T_0$ is true, Jovians do not exist, and the reference class contains only humans, which makes us typical. Typicality suggests penalizing theories according to which we are atypical, so, by the assumptions made, it suggests discarding $T_J$ even if our empirical evidence otherwise happens to be neutral between $T_J$ and $T_0$ or even slightly favors $T_J$.

The unattractive preference for $T_0$ disappears if we do not include the Jovians in the reference class. But, assuming that the hypothesized Jovians are sentient and intelligent, it is unclear on what basis one could justify their exclusion. A well-motivated demarcation criterion for observer reference class membership is clearly needed.

As pointed out by Garriga and Vilenkin \cite{garrigavilenkin}, it is possible to avoid the unattractive systematic preference for $T_0$ over $T_J$ without excluding the Jovians from the reference class by invoking the controversial \textit{self-indication assumption} (SIA) \cite{bostrom_book}. According to the SIA, which assumes a Bayesian framework of theory assessment, we should assign \textit{prior} probabilities in such a way that theories are privileged and/or disfavoured in proportion to the number of observers whose existence they entail. (The SIA is independently advocated by Olum \cite{olum}.) When applied to the humans and Jovians scenario, the SIA precisely cancels the intuitively implausible systematic preference for $T_0$ over $T_J$. However, as pointed out by Bostrom and Cirkovic \cite{cirkovic} and acknowledged by Garriga and Vilenkin, the SIA has highly counterintuitive consequences in other domains, notably the so-called \textit{presumptuous philosopher problem} (\cite{bostrom_book}, p.\ 124). The SIA does in any case not resolve the demarcation problem for observer reference class membership.

In cosmological practice, relative observer numbers are generally estimated by relying on observer \textit{proxies} such as, for example, relative amount of matter clustered in giant galaxies \cite{martelshapiroweinberg} or entropy gradient in the causal diamond \cite{bousso,boussoharnik}. The observer reference class problem manifests itself in the difficulties that arise in the choice of a concrete proxy, which is usually based on assumptions concerning which observers exist according to the theories under consideration that qualify for inclusion in the observer reference class. The fact that the observer reference class problem is relevant to proxy choice underscores how pressing it is: as pointed out by Starkman and Trotta \cite{starkmantrotta}, predictions from multiverse theories for the cosmological constant depend dramatically on the proxy.

The present paper focuses on an approach to solving the observer reference class problem that derives from the work of Srednicki and Hartle \cite{hartlesrednicki_new}. Their approach suggests solving the observer reference class problem by empirically \textit{testing} typicality with respect to various candidate reference classes and, in the end, opting for the competitively most successful reference class. The following section reviews Srednicki and Hartle's approach, the subsequent section criticizes it.

\section{Srednicki and Hartle: the ``\,`framework' framework''}

As we have seen, typicality is based on \textit{self-locating indifference}: the assignment of a uniform probability distribution over observers in the chosen reference class. Following Srednicki and Hartle, I refer to probabilities ascribed to who one might be among those in the reference class as \textit{first-person} probabilities and to probability distributions $\xi$ that ascribe first-person probabilities as \textit{xerographic distributions}. If there are $N$ observers in the reference class, the uniform xerographic distribution, which expresses typicality, is given by $\xi_{ind}(x_i)=1/N$ (where the variable $x_i$ denotes observers in the reference class and the index ``$ind$'' stands for ``indifference''). An example by Bostrom illustrates the appeal of self-locating indifference:
\begin{quote}
	The world consists of a dungeon that has one hundred cells. In each cell there is one prisoner. Ninety of the cells are painted blue on the	outside and the other ten are painted red. Each prisoner is asked to guess whether he is in a blue or a red cell. (Everybody knows all this.) You find yourself in one of the cells. What color should you think it is?---Answer: Blue, with $90\%$ probability. (\cite{bostrom_book}, p.\ 59f.)
\end{quote}
The answer $P^{1p}({\rm My\ cell\ is\ blue})=90\%$ clearly follows from $\xi_{ind}(x_i) = 1/100$, where $i= 1, ...,100$ labels observers by cell number. As pointed out by Bostrom (\cite{bostrom_book}, p.\ 60f.), the distribution $\xi_{ind}$ can be motivated in that it is the only credence function which, when simultaneously adopted by all prisoners, prevents them from a sure loss in a well-chosen hypothetical collective bet against them. This motivation, however, does not address the question of why the observer reference class---encompassing those ascribed non-zero probability by the xerographic distribution---should be chosen to include precisely all prisoners rather than only some of them and/or also some non-prisoners (if such exist).

The framework proposed by Srednicki and Hartle in \cite{hartlesrednicki_new} suggests that fundamental physical theories $T$ be tested alongside xerographic distributions $\xi$ in dyads $(T,\xi)$ called ``frameworks''. (Hence the title of the present section.) Testing frameworks is conveniently modelled along Bayesian lines (see \cite{barnes_new} for a recent systematic defence of using Bayesianism in testing multiverse theories). The starting point is a prior ``third-person'', probability distribution $P(T,\xi)$ over candidate frameworks $(T,\xi)$. Using Bayes' theorem, first-person probabilities $P^{1p}(T,\xi|D_0)$, conditional with respect to our complete background information $D_0$, can be obtained from the third-person probabilities using Bayes' theorem (see Eq.\ 4.1 in \cite{hartlesrednicki_new}):
\begin{eqnarray}
P^{1p}(T,\xi|D_0)= \frac{P^{1p}(D_0|T,\xi)P(T,\xi)}{\sum_{(T,\xi)}P^{1p}(D_0|T,\xi)P(T,\xi)}\,.\label{1-p}
\end{eqnarray}
A further crucial assumption by Srednicki and Hartle is that we can treat the fundamental physical theory $T$ and the xerographic distribution $\xi$ as independent in that the third-person probability distribution over frameworks $(T,\xi)$ factorizes into a contribution $P_{th}$ over theories and a contribution $P_{xd}$ over xerographic distributions:
\begin{eqnarray}
P(T,\xi)= P_{xd}(\xi)\cdot P_{th}(T)\,.\label{factorize}
\end{eqnarray}
By Eq.\ (\ref{factorize}), we can assess the impact of incoming empirical evidence with respect to $P_{xd}$ and $P_{th}$ separately. Notably, this has the consequence that we can ``compete'' different xerographic distributions against each other for the same theory $T$. Since typicality with respect to any specific reference classes can be expressed as indifference $x_{ind}(x_i)=1/N$ over the observer reference class, this ``competition'' promises an empirical solution to the observer reference class problem with the victorious reference class selected as the one to be used in future empirical tests of multiverse (and other cosmological) theories.

\section{Criticizing the ``\,`framework' framework''}

The xerographic distribution $\xi(x_i)$ is itself a probability distribution, so the probabilities ascribed to xerographic distributions by $P_{xd}(\xi)$ are second-order. There is no difficulty in principle with second-order probabilities. For example, hypotheses $H_\theta$ concerning the unknown bias $\theta$ of some (presumably biased) coin can be individuated in terms of their probability ascriptions to the various possible sequences of outcomes. Accordingly, probabilities $P(H_\theta)$ defined over such hypotheses concerning coin bias are second-order. They are obviously well-defined and can be tested by repeated coin tossing and evaluating the toss outcomes.

However, there is an important difference between hypotheses $H_\theta$ concerning coin bias and xerographic distributions $\xi$: while there supposedly exists some actual---perhaps unknown---bias $\theta$ of the coin (defined in terms of long-term outcome frequencies), so that precisely one $H_\theta$ is true, there is no ``true'' xerographic distribution $\xi$, except in the uninteresting and trivial sense in which each observer $x_j$ has this role played by their characteristic function $\chi_j(x_i)=\delta_{i,j}$.

This fundamental difference between hypotheses concerning unknown coin biases and xerographic distributions becomes relevant when it comes to updating procedures with respect to $\xi$ and $H_\theta$, respectively. Whereas the evidential impact of data concerning coin toss outcomes on probability assignments over the hypotheses $H_{\theta}$ can be conveniently modelled by Bayesian updating of $P(H_\theta)$, it turns out that the impact of self-locating information is adequately modelled only by updating the xerographic distribution $\xi$ \textit{itself} rather than by updating any probability distribution $P_{xd}(\xi)$ over xerographic distributions.

To see this, consider again Bostrom's dungeon, which hosts 90 prisoners in blue cells and 10 prisoners in red cells. Suppose that a prisoner somehow finds out that she is in a blue cell. According to the approach suggested by Srednicki and Hartle, her rational posterior first-person probability $P^{>,1p}(\xi)$ after finding this out, which is given by her conditional first-person prior probability $P^{1p}(\xi|{\rm My\ cell\ is\ blue})$, evaluated at fixed theory $T$ for which $P_{th}(T)=1$, is obtained from Eq.\ (\ref{1-p}) as
\begin{eqnarray}
P^{1p}(\xi|{\rm My\ cell\ is\ blue})= \frac{P^{1p}({\rm My\ cell\ is\ blue}|\xi)P_{xd}(\xi)}{P^{1p}({\rm My\ cell\ is\ blue})}\,.\label{1-p_red}
\end{eqnarray}
Now suppose that the prior $P_{xd}(\xi)$ over xerographic distributions ascribes a non-zero probability to at least one xerographic distribution $\xi_0$ that has non-vanishing support in both blue and red cells. For example, suppose that the prior $P_{xd}(\xi_{ind})$ assigned to $\xi_{ind}(x_i) = 1/100$ is non-zero (possibly $1$).

By assumption, $\xi_0$ ascribes non-zero probability to being in a blue cell, i.e. \\$P^{1p}({\rm My\ cell\ is\ blue}|\xi_0)$ is non-zero. By Eq.\ (\ref{1-p_red}), $P^{1p}(\xi_0|{\rm My\ cell\ is\ blue})$ must be non-zero as well. This is problematic, however: by assumption, $\xi_0$ ascribes non-zero probability to being an observer in a red cell, and this conflicts with one's determinate knowledge that one's cell colour is blue. Updating in accordance with Eq.\ (\ref{1-p_red}) thus leads to an awkward situation where one is, on the one hand, completely certain that one is in a blue cell while simultaneously entertaining the possibility that one is in a red cell \footnote{No analogous difficulty arises with respect to hypotheses $H_\theta$ concerning coin toss bias. Here the posterior $P^>(H_\theta)=P(H_\theta|D)$, where $D$ are outcome data concerning observed tosses, is given by
\[
P(H_\theta|D)= \frac{P(D|H_\theta)P(H_\theta)}{P(D)}\,
\]
This is unproblematic because being certain about $D$ is not in tension with assigning a non-zero (posterior) probability to $H_\theta$. Whether $H_\theta$ obtains concerns the limiting long-run frequency behaviour of the coin, assumed to be a determinate matter, which is not in any possible conflict with the outcome data $D$.}.

The problem is not confined to simple examples like Bostrom's dungeon but appears in applications of Srednicki and Hartle's framework to actual cosmological problems. For example, assume that we want to account for the observed value of the cosmological constant $\Lambda$ and assign a prior $P_{xd}(\xi)$ that is non-zero (perhaps $1$) for the uniform xerographic distribution $\xi_{ind}$ over observers who witness the full anthropically allowed range $\Delta_\Lambda$ across universes. Having measured $\Lambda$ in our own universe and finding it within some finite proper subrange $\delta_{\Lambda}\subset\Delta_\Lambda$, the posterior first-person probability $P^{>,1p}(\xi_{ind})=P^{1p}(\xi_{ind}|\Lambda\in\delta_{\Lambda})$ assigned to $\xi_{ind}$ is non-zero by Bayes' theorem because $\xi_{ind}$ is non-zero over the members of $\delta_{\Lambda}$. Assigning a non-zero posterior to $\xi_{ind}$ in that situation is incoherent, however, because $\xi_{ind}$ ascribes non-zero probability to $\Lambda$ lying in $\Delta_\lambda$ but \textit{outside} $\delta_{\Lambda}$, contrary to our knowledge from measurements of $\Lambda$ according to which $\Lambda\in\delta_{\Lambda}$.

The adequate procedure for taking into account the impact of self-locating information like ``My cell is blue'' or ``I'm in a universe with $\Lambda\in\delta_{\Lambda}$'' does not seem to be Bayesian updating of any second-order probability distribution $P_{xd}(\xi)$. A much simpler and better option is to directly update some prior xerographic distribution $\xi_{prior}$ \textit{itself}. (If there are various candidate priors $\xi_k$ between which it is difficult to decide, one can use a weighted average $\xi=\sum_kw_k\xi_k$, where the $w_k$ can be chosen such as to correspond as the $P_{xd}(\xi_k)$ in Srednicki and Hartle's scheme.)

This option immediately delivers the intuitively correct and plausible verdict for the applications discussed. For example, if in Bostrom's dungeon a prisoner starts with the uniform prior $\xi_{ind}(x_i)=1/100$ over all prisoners, ordinary Bayesian conditioning with respect to ``My cell is blue'' yields the attractive posterior $\xi^>(x_i)=\xi_{ind}(x_i|{\rm My\ cell\ is\ blue})$ given by the conditional prior
\begin{eqnarray}
\xi_{ind}(x_i|{\rm My\ cell\ is\ blue})= \frac{\xi_{ind}({\rm My\ cell\ is\ blue}|x_i)\cdot \xi_{ind}(x_i)}{\xi_{ind}({\rm My\ cell\ is\ blue})}\nonumber\\
=\begin{cases}\frac{1\cdot 1/100}{9/10}\ {\rm if\ }x_i\ {\rm is\ a\ blue-cell\ observer}\\\frac{0\cdot 1/100}{9/10}\ {\rm otherwise}\end{cases}\nonumber\\
= \begin{cases}1/90\ {\rm if\ }x_i\ {\rm is\ a\ blue-cell\ observer}\\0\ {\rm otherwise}\end{cases}\,.\label{xero_update}
\end{eqnarray}
In this updating scheme, unlike in the one that derives from the framework proposed by Srednicki and Hartle, typicality as encoded in $\xi_{ind}$ is not empirically ``tested'' in that it is not treated as a hypothesis in analogy to some theory $T$ that one treats as confirmed or disconfirmed by evidence. Prima facie it seems to be used as a primitive starting point for which no further justification is given. Having abandoned the idea of solving the observer reference class problem by testing typicality with respect to candidate reference classes we will see in the following section how paying attention to one's background information $D_0$ allows one to choose the reference class in a non-arbitrary way.

\section{The background information constraint in reference class choice}

Garriga and Vilenkin suggest that the observer reference class problem can be solved by including in the reference class precisely those observers with ``identical information content'' as oneself \cite{garrigavilenkin}. According to them, in other words, given an observer's full (first- \textit{and} third-person) background information (``information content'') $D_0$, the appropriate reference class to use is the one that contains precisely those whose background information is $D_0$.

This suggestion is helpful but delivers overly restrictive verdicts in some applications. Consider again Bostrom's dungeon scenario: one would expect that prisoners in different cells differ in memories and/or current states of knowledge, but this will not per se make it irrational for them to assign the uniform distribution $\xi(x_i)=1/100$ over prisoners. What \textit{would} make it irrational is if their background information $D_0$ allowed them to rule out being some of the 100 prisoners. This simple idea is encoded in the following \textit{background information constraint} (``BIC'') on the observer reference class, which the present paper suggests as a general principle of observer reference class choice in anthropic reasoning:
\begin{quote}
	(BIC) Given background information $D_0$, include in the observer reference class precisely those observers who you possibly \textit{could be} in view of $D_0$.
\end{quote}
Typicality is only attractive if the background information $D_0$ is evidentially \textit{neutral} with respect to who one might be among those who one possibly \textit{could be}, given $D_0$. If this neutrality constraint is not fulfilled, a non-uniform xerographic distribution should be used in place of the uniform one, still based on the background information constraint BIC. In what follows it will be assumed that the neutrality constraint is fulfilled.

Who among all observers that exist according to some theory $T$ at issue one could possibly be, given one's background information $D_0$, depends in general on the theory $T$. Different fundamental theories $T_1$ and $T_2$ will usually differ on which observers exist and on how many of them one could possibly be, given that $D_0$ is one's background information. Accordingly, xerographic distributions associated with $T_1$ and $T_2$ will in general differ as well. We must therefore assign an index for the theory $T$ to the xerographic distribution, so that it becomes $\xi_T(x_i|D_0)$.

By the definition of conditional probability, the first-person probability $P^{1p}(T,x_i|D_0)$ can be written in product form as 
\begin{eqnarray}
P^{1p}(T,x_i|D_0)= P^{1p}(x_i|T,D_0)\cdot P^{1p}(T|D_0)\,,\label{factorize1}
\end{eqnarray}
which replaces Srednicki and Hartle's factorization assumption Eq.\ (\ref{factorize}).

The first factor on the right hand side of Eq.\ (\ref{factorize1}) specifies the probability of being $x_i$, assuming $T$'s truth and one's background information $D_0$. This makes it plainly equivalent with the xerographic distribution $\xi_T(x_i|D_0)$, which also specifies the probability of being $x_i$, given $D_0$, if $T$ holds. The second factor on the right hand side of Eq.\ (\ref{factorize1}) applies only to the ``third-person'' fact of which theory $T$ holds. One can identify it with the third-person probability distribution $P_{th}(T|D_0)$ over theories. In virtue of these identities Eq.\ (\ref{factorize1}) becomes
\begin{eqnarray}
P^{1p}(T,x_i|D_0)= \xi_T(x_i|D_0)\cdot P_{th}(T|D_0)\,.\label{factorize2}
\end{eqnarray}
The following section illustrates how BIC, using Eq.\ (\ref{factorize2}), solves the observer reference class problem by applying it to the humans and Jovians scenario.

\section{Humans and Jovians again}

Let us recapitulate why we should expect that there are likely no Jovians if we include the Jovians in our observer reference class. The indifference-expressing xerographic distribution in this case is $\xi_{T_J,ind}(x_i|D_0)=1/(N_J+N_H)$ for $T_J$ and $\xi_{T_0,ind}(x_i|D_0)=1/N_H$ for $T_0$, where $N_J$ is the number of Jovians according to $T_J$ and $N_H$ the (known) number of humans according to both $T_J$ and $T_0$. 

From Eq.\ (\ref{factorize2}) we obtain the ratio of the first-person probabilities $P^{1p}(T_J,{\rm I\ am\ human}|D_0)$ and $P^{1p}(T_0,{\rm I\ am\ human}|D_0)$ as
\begin{eqnarray}
\frac{P^{1p}(T_J,{\rm I\ am\ human}|D_0)}{P^{1p}(T_0,{\rm I\ am\ human}|D_0)}= \frac{N_H}{N_J+N_H}\cdot\frac{P_{th}(T_J|D_0)}{P_{th}(T_0|D_0)}\,.\label{compare}
\end{eqnarray}
By assumption, $N_H/(N_J+N_H)\ll1$ because $N_H\ll N_J$. As a consequence, unless the third-person probabilities $P_{th}(T_J|D_0)$ and $P_{th}(T_0|D_0)$ happen to strongly favor $T_J$ over $T_0$, the first-person probabilities $P^{1p}(T_{J}|D_0,{\rm I\ am\ human})$ and $P^{1p}(T_{0}|D_0,{\rm I\ am\ human})$ turn out to strongly favor $T_0$ over $T_J$. Taking into account the datum that we are humans thus supports the null hypothesis $T_0$ according to which there are no Jovians over the alternative hypothesis $T_J$ according to which there are Jovians.

But should we include the Jovians in the observer reference class according to BIC? To answer this question we need to determine whether we possibly could be Jovians, given our background information $D_0$. Now, our \textit{actual} background information $D_0$, which includes all (scientific and everyday) data that we currently happen to have, plainly includes the fact that we are human and live on Earth. Plausibly, it not only rules out that we are Jovians but even specifies precisely who we are among humans, identifying each of us in terms of name, date of birth, place of birth etc. Our actual background information thus effectively narrows down who we are to some specific human observer $x_{j}$. So, using our actual background information as $D_0$ we obtain from Eq.\ (\ref{factorize2})
\begin{eqnarray}
	\frac{P^{1p}(T_J,{\rm I\ am\ human}|D_0)}{P^{1p}(T_0,{\rm I\ am\ human}|D_0)}=\frac{P_{th}(T_J|D_0)}{P_{th}(T_0|D_0)}\,,
\end{eqnarray}
which yields no preference in first-person probabilities for $T_0$ over $T_J$ when taking into account one's being human.

But perhaps we should consider operating on the basis of background information $D_0$ which does \textit{not} include the fact that we are human. Indeed, it seems natural to \textit{abstract} from data whose impact we are trying to assess---in this case, that we are human---and consider that impact based on background information $D_0$ that does not include these data. And if some hypothetical amount of background information $D_0$ has been obtained by abstracting from our being human, it may well be compatible with its bearer being Jovian. Then BIC would dictate using a reference class that indeed does have Jovians as members. For such background information $D_0$ the argument centred around Eq.\ (\ref{compare}) can be run and the systematic preference for $T_0$ apparently vindicated.

However, while abstracting from bits of our actual background information to arrive at a suitable $D_0$ is in principle possible, the actual, practical outcome of this abstraction process seems highly dubious and unclear. We may ask ourselves, for example, whether hypothetical background information $D_0$ that is compatible with us being Jovians should plausibly include data gathered at particle accelerators built and operated by us humans and, if some, which of those data. The answer to this question has the potential to matter a lot with respect to who exactly will end up included in the reference class.

Moreover, since any candidate $D_0$ arrived at by abstracting from the fact that we are human is so radically impoverished with respect to our \textit{actual} background information, it is difficult to estimate which third-person probability assignments $P_{th}(T_J|D_0)$ would be rational to assign. Our ordinary plausibility verdicts are probably no reliable guide if the hypothesized background information $D_0$ differs so massively from our actual state of knowledge. But unless we have a prima facie reasonable case for assigning values $P_{th}(T_J|D_0)$ and $P_{th}(T_0|D_0)$ of specific orders of magnitude, we have, in the light of Eq.\ (\ref{factorize2}), no strong case for first-person probabilities $P^{1p}(T_{J}|B,{\rm I\ am\ human})$ and $P^{1p}(T_{0}|B,{\rm I\ am\ human})$ that strongly favor $T_0$ over $T_J$. To conclude, BIC solves the observer reference class problem in such a way that, when applied to the humans and Jovians scenario, there appears to be no robust case for any unattractive systematic preference of $T_0$ over $T_J$.

\section{Accounting for measured parameters using multiverse theories}

Which background information $D_0$ and, derivatively, which reference class of observers should we choose in attempts to account for the measured values of fundamental parameters in terms of multiverse theories? The most obvious choice would be our full actual background information and, correspondingly, the reference class of all observers who we possibly could be, given our actual background information and the theory $T$ that we happen to consider. However, if we want to account for the value of some already measured parameter, this is not an option because it would prevent us from treating that value as a non-trivial prediction of $T$. For example, if we want to account for the measured value of the cosmological constant using the landscape multiverse scenario, we must use a reference class of observers who find different values for that constant and then determine whether the value that we find is a typical one among those found by observers in that reference class. 

There is probably no single best recipe for singling out any specific proper part $D_0$ of our actual background information as ideally suited for this task. The following two desiderata, however, seem reasonable to request and may serve as rough guidelines:
\begin{itemize}
	\item First, it should be practically feasible to determine unambiguously which and how many observers one could possibly be, given the chosen background information $D_0$, for all theories $T$ that one wishes to test.
	\item Second, it should be practically possible to reasonably motivate one's assigned third-person probabilities $P_{th}(T|D_0)$ for the chosen $D_0$ and all theories $T$ that one wishes to test.
\end{itemize}
Figuring out which observers exist according to some given multiverse theory $T$ and how many of them one could possibly be, given some chosen background information $D_0$, is in general a Gargantuan challenge. Typically, it will require predicting and understanding complex and emergent phenomena in universes that have radically different effective (higher-level) physical laws and parameters than our own. But despite these difficulties, accepting the challenge is without alternatives, notably if one wishes to check the quality of observer proxies and, derivatively, the predictions based on them. Thus, even though the rule BIC in principle allows to uniquely identify the appropriate observer reference class to be used in applications of typicality, testing concrete multiverse theories on its basis will likely remain an arduous task.

\section*{Acknowledgement}
Work on this article was supported by the Netherlands Organization for Scientific Research (NWO), Veni grant 275-20-065.


\begin{thebibliography}{}
	\bibitem{boussopolchinski} R. Bousso and J. Polchinski, J. High Energy Phys. \textbf{06}, 006 (2000).
	\bibitem{kachru} S. Kachru, R. Kallosh, A. Linde, and S. P. Trivedi, Phys, Rev. D \textbf{68}, 046005 (2003).
	\bibitem{susskind} L. Susskind, \textit{The Cosmic Landscape: String Theory and the Illusion of Intelligent Design}	(New York: Back Bay Books, 2005).
	\bibitem{vilenkin} A. Vilenkin, Phys. Rev. Lett. \textbf{74}, 846 (1995).
	\bibitem{bostrom_book} N. Bostrom, \textit{Anthropic Bias: Observation Selection Effects in Science and Philosophy} (New York: Routledge, 2002).
	\bibitem{elga_ind} A. Elga, Philosophy and Phenomenological Research \textbf{69}, 383 (2004).
	\bibitem{aguirretegmark} A. Aguirre and M. Tegmark, J. Cosm. and Astrop. Phys. \textbf{501}, 003 (2005).
	\bibitem{weinstein} S. Weinstein, Class. and Quantum Grav. \textbf{23}, 4231 (2006).
	\bibitem{hartlesrednicki} J. B. Hartle and M. Srednicki, Phys. Rev. D \textbf{75}, 123523 (2007).
	\bibitem{azhar} F. Azhar, Class. Quantum Gravity \textit{31}, 035005 (2014).
	\bibitem{hartlesrednicki_new} M. Srednicki and J. B. Hartle, Phys. Rev. D \textbf{81}, 123524 (2010).
	\bibitem{weinberg} S. Weinberg, Phys. Rev. Lett. \textbf{59}, 2607 (1987).
	\bibitem{boussoharnik} R. Bousso, R. Harnik, G. D. Kribs, and G. Perez, Phys. Rev. D \textbf{76}, 043513 (2007).
	\bibitem{desimone} A. De Simone, A. H. Guth, M. P. Salem, and A. Vilenkin, Physical Review D \textbf{78}, 063520 (2008).
	\bibitem{garrigavilenkin} J. Garriga and A. Vilenkin, Phys. Rev. D \textbf{77}, 043526 (2008).
	\bibitem{olum} K. Olum, The Philosophical Quarterly \textbf{52}, 164 (2002).
	\bibitem{cirkovic} N. Bostrom and M. M. \'{C}irkovi\'{c}, The Philosophical Quarterly \textbf{53}, 83 (2003).
	\bibitem{martelshapiroweinberg} H. Martel, P. R. Shapiro, and S. Weinberg, Astrophys. Journal \textit{492}, 29 (1998).
	\bibitem{bousso} R. Bousso, Phys. Rev. Lett. \textbf{97}, 191302 (2006).
	\bibitem{starkmantrotta} G. D. Starkman and R. Trotta, Phys. Rev. Lett. \textbf{97}, 201301 (2006).
	\bibitem{barnes_new} L. A. Barnes, in \textit{The Philosophy of Cosmology}, edited by K. Chamcham, J. Barrow, S. Saunders, and J. Silk (Cambridge: Cambridge University Press, 2017) p. 447.
	\bibitem{carroll_book} S. Carroll, \textit{From Eternity to Here} (New York and London: Plume, 2010).
\end{thebibliography}
\end{document}